**Chapter Number**

# Employing Soft X–rays in Experimental Astrochemistry

Sergio Pilling and Diana P. P. Andrade
*Inst. de Pesquisa & Desenvolvimento, Universidade do Vale do Paraíba (UNIVAP),*
*São Jose dos Campos*
*Brazil*

## 1. Introduction

Young stars and protostars are hot enough to produce a strong ionizing field with the maximum at ultraviolet and soft X-rays (Koyama et al. 1996). These objects are usually associated with regions in which several stars are borning together, called star-forming regions, inside of molecular clouds. Examples of such regions are Sgr B2, Orion KL, and W51. In these star-forming regions, the presence of widespread UV and X-ray fields could trigger the formation of photodissociation regions (PDRs). X-ray photons are capable of traversing large column densities of gas before being absorbed, which produce X-ray-dominated regions (XDRs) where the interface between the ionized gas and the self-shielded neutral layers could influence the selective heating of the molecular gas (Goicoechea et al. 2004). The typical energy flux in XDRs inside molecular clouds is about ~10 erg cm$^{-2}$s$^{-1}$, which decreases as the visual extinction increases within the clouds. The complexity of these regions may allow a combination of different scenarios and excitation/ionization mechanisms to coexist.

Figure 1 shows a typical photodissociation region (Gaseous Pillars from M16 nebulae) sculpted by radiation pressure of young stars inside molecular clouds obtained by the hubble space telescope.

The photochemistry induced by soft X-rays on molecular gas produces excited states (both valence and core levels), radicals, ions (cations and anions), and excited ions, as well as promotes the release of fast electrons. Such electrons can also trigger further excitation/ionization processes. In solid phase (such as interstellar ices), the main effects of soft X-rays are local heating, molecular excitation/ionization, molecular bonds rupture (photodissociation), and photodessorption (releasing of ion or neutral species from the frozen phase to the gas phase). In addition, interstellar grain charging can occur due to the photoelectron effect on the grain surfaces induced by soft X-rays.

In both gas and solid phases, the presence of molecules in excited states, radicals, and ions (cations and anions) is highly enhanced due to soft X-rays in comparison with UV photons. Those species react with each other and with neutral gas, enhancing the molecular complexity of the region.

Due to the high absorbance of soft X-rays by the Earth's atmosphere, most of the information of extraterrestrial X-rays sources comes from space telescopes, such as, ROSAT, Chandra, XMM-Newton, Beppo-SAX, Swift, ASCA, and others.

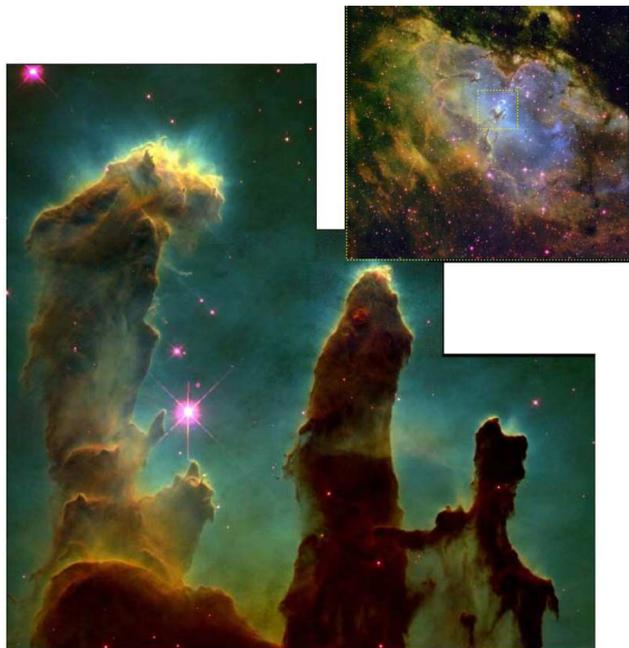

Fig. 1. Photodissociation region known as Gaseous Pillars associated with recent star forming region inside M16 nebulea (Figure inset). The fingerlike structures are the remnants of the molecular clouds which were sculpted by radiation pressure of young stars. (From Hubble Space Telescope 2003)

## 2. Soft X-rays in experimental astrochemistry

Soft X-rays have been used to simulate the interaction between stellar photons and matter in different astrophysical scenarios, including both gas-phase and solid-phase (ices). For these experiments, synchrotron light sources have been employed because of their high intensity and wide wavelength range (coverage). Most of the molecular species investigated were detected in different astrophysical environments (e.g. interstellar medium, comets). In this chapter, we report several experiments performed at the Brazilian Synchrotron Light Laboratory (LNLS) in Campinas, São Paulo, Brazil.

### 2.1 Gas-phase experiments

Soft X-ray photons ($\sim 10^{12}$ photons/s) from a toroidal grating monochromator (TGM) beamline (100–310 eV) perpendicularly intersected the gas sample inside a high vacuum chamber. The gas needle was kept at ground potential. The photon flux was recorded by a light sensitive diode. Several species were investigated in the gas-phase mode, including formic acetic, formic acid, methanol, ethanol, methyl formate, methylamine, acetone, and benzene, as well as some amino acids and nucleobases. These species have been detected or predicted to be present in interstellar medium.

Conventional time-of-flight mass spectra (TOF-MS) were obtained using the correlation between one Photoelectron and a Photoion Coincidence (PEPICO). Details of Time of flight mass spectrometry are given elsewhere (e.g. Boechat-Roberty et al. 2005, Pilling et al. 2006, Pilling 2006). A schematic diagram of the experimental setup employed in gas-phase experiments is shown in Fig. 2a. Figure 2b is a photograph of the equipment coupled to the soft X-ray beamline of LNLS.

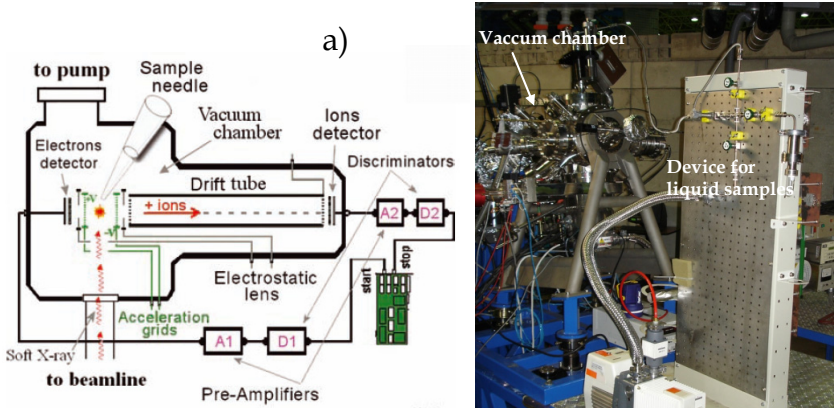

Fig. 2. a) Schematic diagram of the experimental setup employed in gas-phase experiments. b) Photography of equipment employed at the soft X-ray beamline of LNLS.

Briefly, the ionized recoil fragments produced by the interaction with the photon beam were accelerated by a two-stage electric field and detected by two microchannel plate detectors in a chevron configuration, after mass to-charge (m/q) analysis by a time-of-flight mass spectrometer. They produced up to three stop signals for a time-to-digital converter (TDC) started by the signal from one of the electrons accelerated in the opposite direction and recorded without energy analysis by two micro-channel plate detectors. The integrated area of the given peak (ion) in mass spectra divided by the total area (total collected ions) times 100 percent is called the branching ratio (BR). The measurements were done at room temperature. Figure 3 illustrates the particle tracks inside the TOF spectrometers and the potentials employed in these experiments. The simulations were performed using the particle track software SIMION®.

Considering that the kinetic energy of the ionic fragments inside the interaction region is negligible compared with the extraction electron field, the equations for the time of flight of ions inside the spectrometer can be given by

$$\begin{aligned} t_{voo_i} &= t_d + t_L + t_D \\ t_{voo_i} &= \frac{\sqrt{\frac{q}{m}E_d d}}{\frac{q}{m}E_d} + \frac{\sqrt{\frac{q}{m}(E_d d + E_L L)} - \sqrt{\frac{q}{m}E_d d}}{\frac{q}{m}E_L} + \frac{D}{\sqrt{\frac{q}{m}(E_d d + E_L L)}} \end{aligned} \quad (1)$$

and, for the electrons, by the equation

$$\begin{aligned} t_{voo_e} &= t'_d + t_{L_1} + t_{L_2} \\ t_{voo_e} &= \frac{\sqrt{\frac{|q_e|}{m_e}E_d d}}{\frac{|q_e|}{m_e}E_d} + \frac{\sqrt{\frac{|q_e|}{m_e}(E_d d + E_{L_1} L_1)} - \sqrt{\frac{|q_e|}{m_e}E_d d}}{\frac{|q_e|}{m_e}E_{L_1}} + \frac{\sqrt{\frac{|q_e|}{m_e}(E_d d + E_{L_1} L_1 + E_{L_2} L_2)} - \sqrt{\frac{|q_e|}{m_e}(E_d d + E_{L_1} L_1)}}{\frac{|q_e|}{m_e}E_{L_2}} \end{aligned} \quad (2)$$

where L2, L1, d (extraction region), L, and D (drift tube) represent the length of specific regions inside the TOF spectrometer (see Fig. 3).

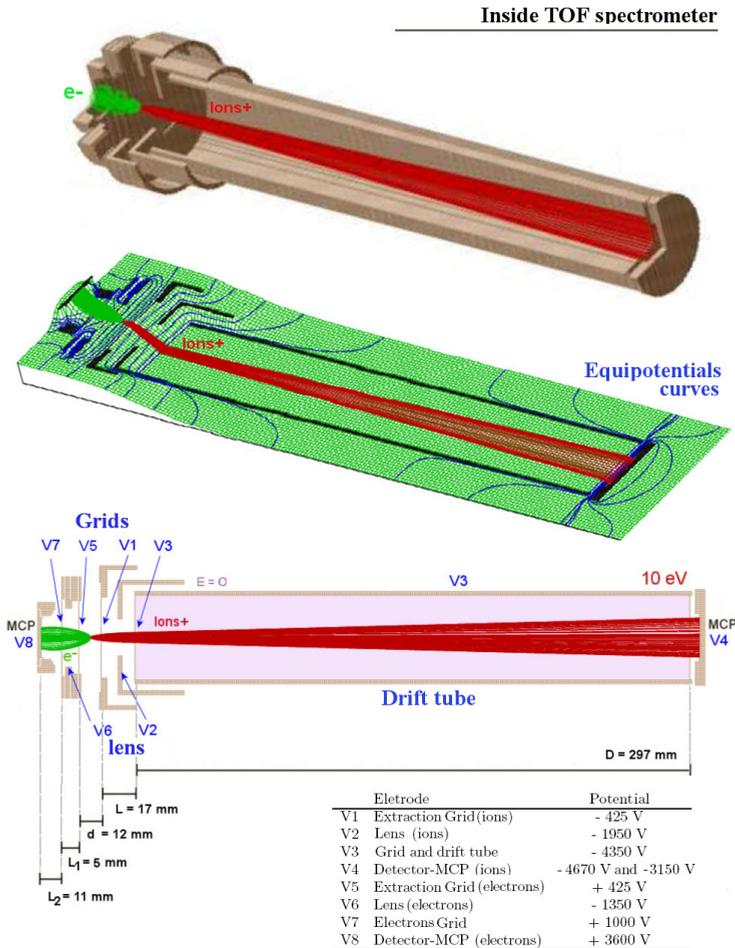

Fig. 3. Particle tracks inside the TOF spectrometers and the potentials employed in the experiments. The simulations were performed using the particle track software SIMION®.

However, due to the employed coincidence technique, the ion's time of flight is not necessarily the time of their detections, $t_{det}$, because one photoelectron, produced during the photoionization process, must reach the detector to start counting the time of the ion´s flight ($t_i \gg t_e \neq 0$). By knowing the electric field inside the TOF spectrometer (Ed = 850/0.012 = 70833 V/m, EL = 3950/0.0017 = 230882 V/m, EL1 = 575/0.005 = 115000 V/m and EL2= 2600/0.011 = 236364 V/m), we can estimate the time of flight of protons ~ 415 ns and of electrons ~1 ns to reach their respective detectors.

Strictly, the time of detection, $t_{det}$, of a given ion is the difference between the ion's time of flight, $t_{voo_i}$, and the time of flight of its respective photoelectron, $t_{voo_e}$,

$$t_{det} = t_{voo_i} - t_{voo_e} \tag{3}$$

Therefore, employing this condition in the Eq. [1] and [2] we obtain

$$t_{det} = \frac{a'}{\sqrt{\frac{q}{m}}} + b' \tag{4}$$

where 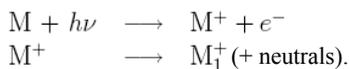 and $b' = -t_{voo_e}$

And finally, we obtain the quadratic relation between the m/q ratio (mass/charge) and the time of flight of ions:

$$\frac{m}{q} = \left(\frac{t_{det} - b'}{a'}\right)^2 = at_{det}^2 + bt_{det} + c \tag{5}$$

where $a = 1/a'^2$, $b = -2b'/a'^2$ and $c = b'^2$ are constants to be determined during the calibration of TOF spectrometer by employing time of flight of fragments with known m/q (eg. Ar+ at 40 u, Ar++ at 20u and, Ar+++ at 13.3u). From Eq. [5] we observe that the spectrometer's resolution is related to the m/q ratio, decreasing for higher values of m/q.

**2.1.1 Single coincidence mass spectrum (PEPICO)**

The simple configuration of the TOF spectrometer is to produce a single coincidence mass spectrum, or in other words, the coincidence between one photo-electron and one photo-ion (PEPICO). The typical reaction scene that occurs inside the ionization region is illustrated by

$$\begin{aligned} M + h\nu &\longrightarrow M^+ + e^- \\ M^+ &\longrightarrow M_1^+ \text{ (+ neutrals)}. \end{aligned}$$

After the absorption of a soft X-ray photon, the molecule releases one electron and becomes positively charged. Following this stage, the excited ion can also dissociate to produce ionic and neutral fragments. Each molecule has its specific pattern of fragment production which depends also on the incoming photon and the molecular orbital configuration.

Depending on the incoming soft X-ray photon, the ejected electron is from inner valence orbitals (less energetic photons) or core orbitals (more energetic photons). The absorption cross sections decrease as a function of photon energy, however, at some specific energies, the absorption cross sections increase abruptly. These energies are called resonances and are associated with the exact energies of electrons to be released from the molecular orbitals. Figure 4 shows a schematic view of the absorption probability, as a function of photon energy, from visible to soft X-rays. The absorption probability is highest at UV wavelengths and decreases to the higher energies. At resonance energies, the absorption probability also increases. These resonances represent the exact energies to promote electrons from valence or inner-shell orbital to unoccupied orbitals (photoexcitation) or to the continuum (photoionization).

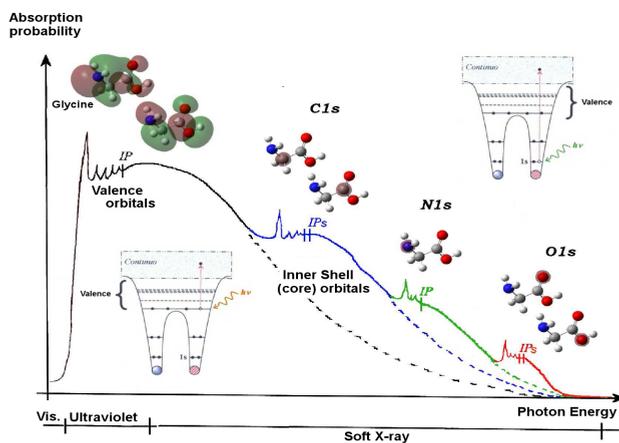

Fig. 4. Schematic view of absorption probability as a function of photon energy for polyatomic molecule (e.g. glycine). IP indicates the ionization potential of a given orbital.

Two examples of PEPICO spectra of water fragments produced by soft X-rays are given in Fig 5. The left spectrum is the original time-of-flight spectra (counts vs. time in ns). The right spectrum was obtained after calibration employing known fragments. This can be performed employing Eq. 5 together with data from two different situations: i) By employing previous experiment with Ar, or ii) by employing the SIMION® program to simulate time-of-flight of different ions. A comparison between these two spectra shows that the peak resolution changes with mass/charge ratio, being worst for higher mass/charge values.

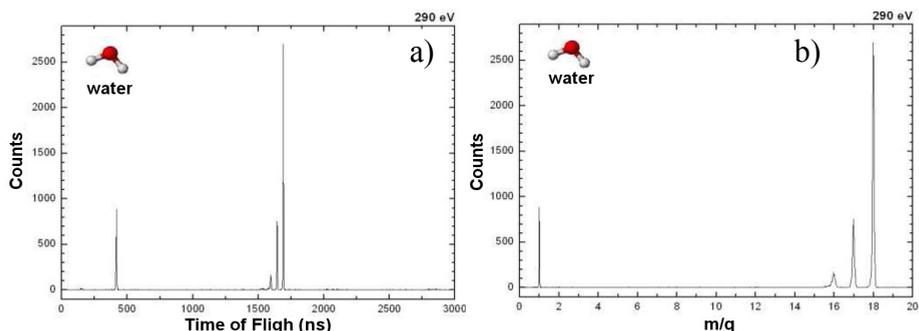

Fig. 5. Sample of time of flight spectra of water fragments from the irradiation with 290 eV before m/q calibration (a) and after m/q calibration (b).

Figure 6 presents some samples of PEPICO mass spectra of studied molecules of astrophysical interest (Pilling et al. 2006; Pilling et al 2011; Ferreira Rodrigues et al. 2008). All spectra were obtained with soft x-rays. The employed soft X-ray energies are shown in the top of each spectrum. Labels close to the peaks indicate possible fragment attributions.

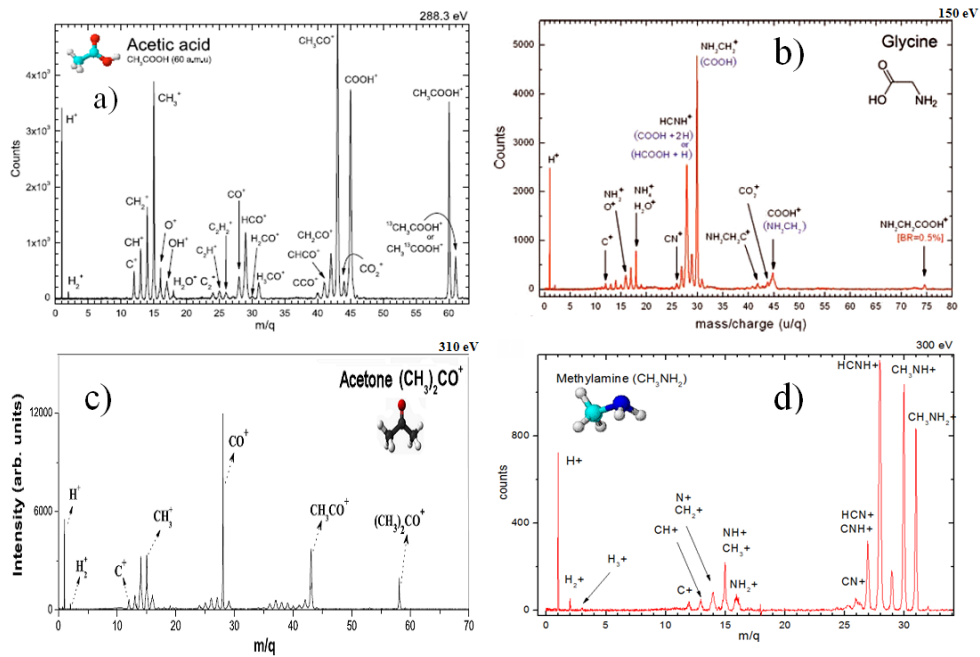

Fig. 6. Time of flight mass spectra (PEPICO) of the fragments produced from the impact of soft X-rays with different gaseous molecules of astrophysical interest. a) acetic acid (Pilling et al 2007) , b) glycine (Pilling et al. 2011), c) Acetone (Ferreira Rodrigues et al. 2008),  and  d) methylamine.

From the analysis of the peaks area, we derived the partial ion yield (PIY) or relative intensities, which correspond to the relative contribution of each cationic fragment in the PEPICO mass spectra. The PIY for each fragment $i$ were obtained by

$$PIY_i = \left( \frac{A_i}{A_t^+} \pm \frac{\sqrt{A_i} + A_i \times ER/100}{A_t^+} \right) \times 100\%$$

where $A_i$ is the area of a fragment peak, $A_t^+$ is the total area of the PEPICO spectrum. The error factor $ER$ (2%) is the estimated error factor due to the data treatment.

Depending on the energy of the incoming photon, a given excitation channel and/or ionization channel will be opened. Figure 7 shows five mass spectra of the fragments produced by the irradiation of the simplest interstellar alcohol (methanol) at different photon energies between 100 and 310 eV (Pilling et al 2007a). The analysis of relative area of individual peaks shows that the area of the molecular ion (parent molecule that lost a electron - $M^+$) peak decreases with the enhancement of photon energy. This suggests that although the absorption probability in soft X-ray range decreases for higher photon energy, the amount of fragmentation produced increases. In other words, the more energy the incoming photon has (in the soft X-ray range), the more the molecule is dissociated in small fragments (atomization).

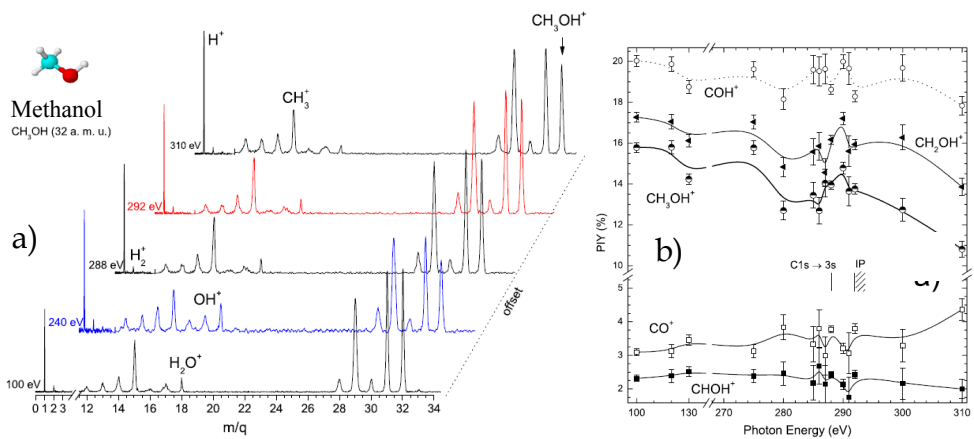

Fig. 7. a) Mass spectra of methanol fragments obtained at different photon energies between 100 and 310 eV. b) Partial ion yield (PIY) of the fragments release by the methanol as a function of photon energy (From Pilling et al 2007a).

Figure 8 shows the mass spectrum (PEPICO) of the fragments of the two simplest interstellar alcohols, methanol and ethanol, irradiated with 292 eV (~42 Angstroms) soft X-rays photons. The arrows indicate the molecular ion in each mass spectrum. From the analysis of the relative area of each peak, we derived the photoionization and photodissociation cross section and also the postproduction cross section of a given fragment. These procedures will be discussed further.

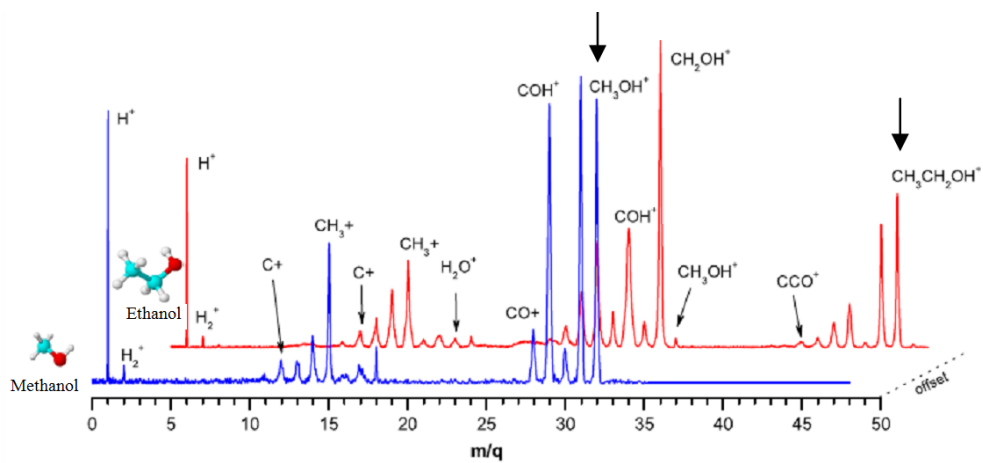

Fig. 8. Mass spectra of the methanol and ethanol fragments produced during irradiation with 292 eV soft X-ray photons.

The mass spectrum of the two simplest carboxylic acids observed in space is given in Figure 9, which presents the fragments produced during the irradiation with 290 eV (~43 Angstroms) soft X-rays photons. The arrows indicate the molecular ion in each mass spectrum. By comparing the relative area of both molecular ions, we observe that formic acid is much more sensitive to this photon energy than acetic acid. Therefore, if the same amount of these molecules is exposed to soft X-rays, after a certain amount of time, the region will be richer in acetic acid than formic acid.

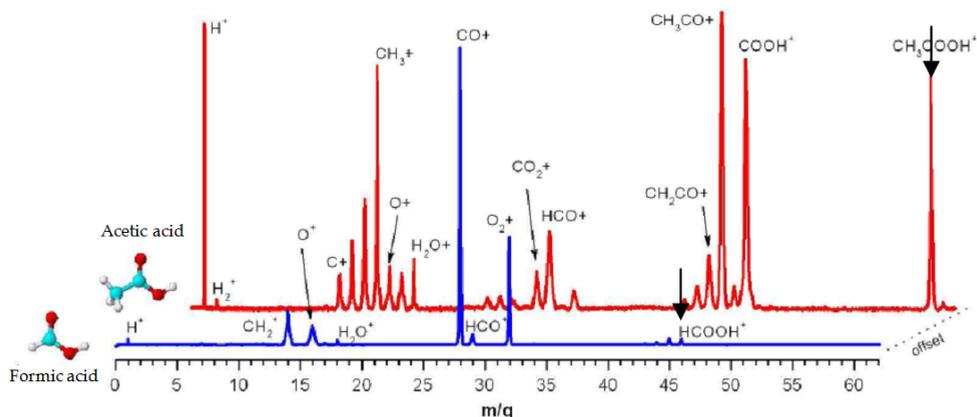

Fig. 9. Mass spectra of the fragments of formic acid and acetic acid produced during the irradiation with 293 eV soft X-ray photons.

**Cross sections**

The absolute cross section values for both photoionization ($\sigma_{ph-i}$) and photodissociation ($\sigma_{ph-d}$) of organic molecules are extremely important as input for molecular abundance models. Sorrell (2001) presented a theoretical model in which biomolecules are formed inside the bulk of icy grain mantles photoprocessed by starlight (ultraviolet and soft X-rays photons). However, the main uncertainty of this equilibrium abundance model comes from the uncertainty of the $\sigma_{ph-d}$ value. Therefore, precisely determining the $\sigma_{ph-d}$ of biomolecules is very important for properly estimating the molecular abundance of those molecules in the interstellar medium that were produced by this mechanism. Moreover, knowing the photon dose $I_0$ and $\sigma_{ph-d}$ values makes it possible to determine the half-life of a given molecule, as discussed by Bernstein et al. (2004).

In order to put the data on an absolute scale, after subtraction of a linear background and false coincidences coming from aborted double and triple ionization (see Simon et al. 1991), we have summed up the contributions of all cationic fragments detected and normalized them to the photoabsorption cross sections available in the literature (e.g. Ishii & Hitchcook 1988). Assuming a negligible fluorescence yield (due to the low carbon atomic number, Chen et al. 1981) and anionic fragment production in the present photon energy range, we assumed that all absorbed photons lead to cation formation. The absolute cross section determination is described elsewhere (Boechat Roberty 2005, Pilling et al. 2006).

Briefly, the non-dissociative single ionization (photoionization) cross-section $\sigma_{ph-i}$ and the dissociative single ionization (photodissociation) cross section $\sigma_{ph-d}$ of studied species can be determined by

$$\sigma_{ph-i} = \sigma^+ \frac{PIY_{M^+}}{100} \qquad (6)$$

and

$$\sigma_{ph-d} = \sigma^+ \left(1 - \frac{PIY_{M^+}}{100}\right) \quad (7)$$

where σ+ is the cross section for single ionized fragments.
From Eq. [6], we can also determine the production cross section of a given cationic fragment, j, from

$$\sigma_{ph-i} = \sigma^+ \frac{PIY_j}{100} \quad (8)$$

Figure 10 presents the ionization and dissociation cross section of simplest interstellar alcohol and carboxylic acids in the soft X-ray energy range around carbon inner shell (C1s).

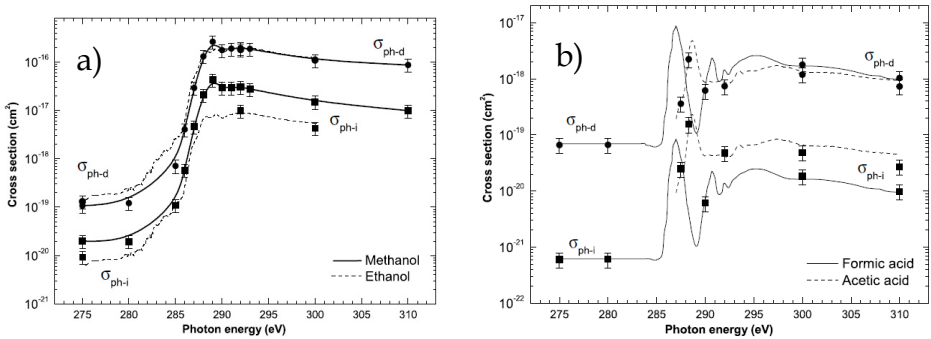

Fig. 10. Ionization and dissociation cross section of a) simples interstellar alcohol and b) carboxylic acids in the soft X-ray energy range around carbon inner shell (C1s).

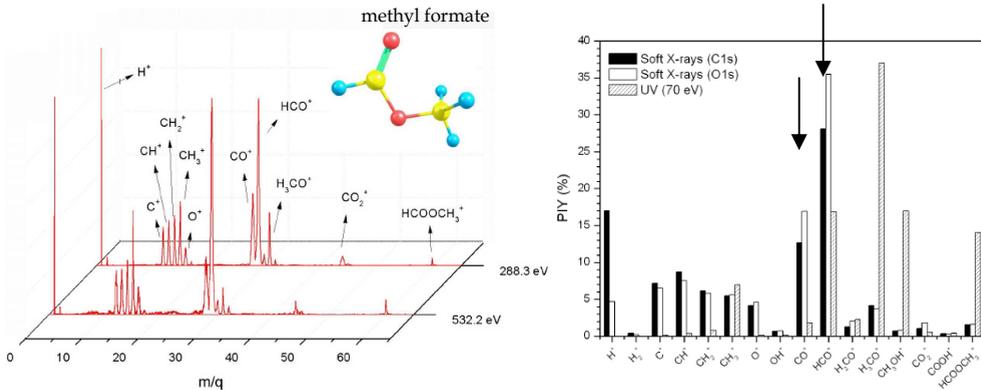

Fig. 11. a) Mass spectra of the methyl formate fragments during irradiation with soft X-ray photons with energies of 288.3 eV (near C1s edge) and 532.2 eV (near O1s edge). b) PIY of methyl formate fragments due to dissociation by soft X-rays (288.3 eV and 532.2 eV) and by UV analogous photon field.

A comparisons between the fragments produced by the dissociation of methyl formate (an isomer of acetic acid) due to soft X-rays with energy around carbon inner shell (C1s) and oxygen inner shell (O1s) is illustrated in Fig. 11a (from Fantuzi et al. 2011). Figure 11b shows a comparison between partial ion yields (PIY) of methyl formate fragments due to dissociation by soft X-rays, at 288.3 eV and 532.2 eV and by 70 eV electrons from the NIST database. The dissociation induced by 70 eV electrons is very similar to the dissociation induced by UV photons

From the PIY analysis at both energies, we observe that for this molecule, the production of CO+ and HCO+ is enhanced when photons at O1s are employed (see arrows). This production is also very enhanced when compared with UV photons. The molecular ion did not have significant changes in this photon energy range.

Another important interstellar molecule studied was benzene ($C_6H_6$). This species may be taken as the basic unit for the polycyclic aromatic hydrocarbons (PAHs), which are believed to play an important role in the chemistry of the interstellar medium (Woods et al. 2003). It may also serve as a precursor molecule to more complex organic compounds, such as amino acids like phenylalanine and tyrosine. Figure 12a shows a comparison of two PEPICO Mass spectra of benzene molecule recorded at (a) UV photons (21.21 eV) and (b) soft X-ray photons (289 eV) (Boechat-Roberty et al. 2009). Figure 12b presents the photoionization and photodissociation cross sections of benzene around carbon C1s edge. The photoabsorption cross-section, $\sigma_{ph-abs}$ (solid line), taken from Hitchcock et al. (1987), is also shown.

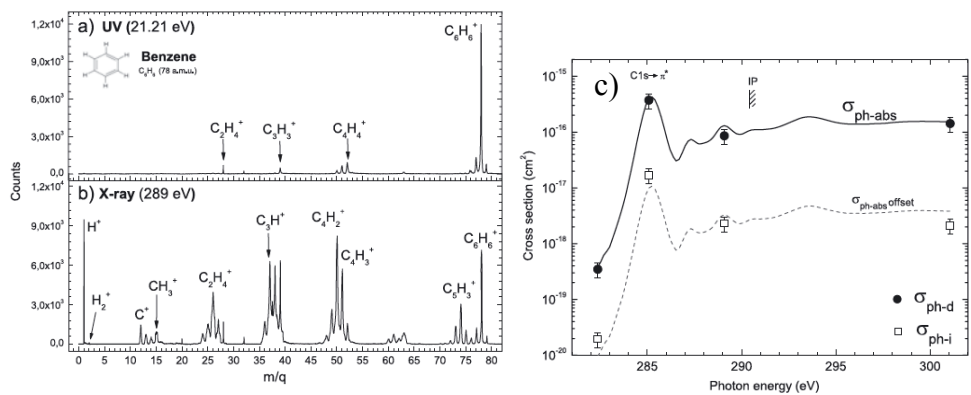

Fig. 12. PEPICO Mass spectra of benzene molecule recorded at (a) UV photons (21.21 eV) and (b) soft X-ray photons (289 eV). c) Photoionization cross section and photodissociation cross section of benzene around C1s edge as a function of photon energy. The dashed line is an off-set of photoabsorption cross-section and is only to guide the eyes (from Boechat-Roberty et al. 2009).

**Kinetic Energy of ionic fragments**

The present time-of flight spectrometer was designed to fulfill the Wiley-McLaren conditions for space focusing (Wiley & McLaren 1955). Within the space focusing conditions, the observed broadening of peaks in spectra is mainly due to kinetic energy release of fragments. A Schematic view of the peak broadening promoted by different orientation of produced fragments with the same initial velocity ($v_o$) at the interact region inside the time-of-flight spectrometer is shown in Figure 13.

Considering that the electric field in the interaction region is uniform, we can determine the released energy in the fragmentation process ($U_0$) from each peak width used by Simon et al. (1991), Hansen et al. (1998), and Santos et al. (2001)

$$U_0 = \left(\frac{qE\Delta t}{2}\right)^2 \frac{1}{2m} \qquad (9)$$

where $q$ is the ion fragment charge, $E$ is the electric field in interaction region, $m$ is the mass of fragment, and $\Delta t$ is the time peak width (FWHM) taken from PEPICO spectra.

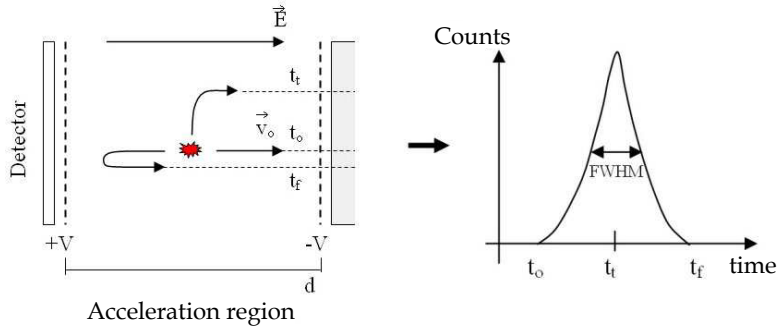

Fig. 13. Schematic view of the peak broadening promoted by different orientation of produced fragments with the same initial velocity ($v_o$) at the interact region inside the time-of-flight spectrometer.

| Fragments | | PIY (%) / $U_0$ (eV) | | | | | | |
|---|---|---|---|---|---|---|---|---|
| m/q | Attribution | 200 eV | 230 eV | 275 eV | 280 eV | 290 eV | 300 eV | 310 eV |
| 1 | H$^+$ | – | 0.71 / 0.24 | 0.19 / 0.24 | 1.33 / 2.17 | 1.32 / 2.17 | 2.36 / 2.87 | 2.83 / 4.80 |
| 7 | CH$_2^{++}$; CO$^{++++}$? | – | 1.57 / 30.5 | 0.86 / 11.0 | 0.72 / 13.8 | 0.47 / 8.84 | 0.52 / 9.97 | 0.64 / 10.0 |
| 8 | O$^{++}$ | – | – | 0.65 / 2.45 | 0.57 / 12.0 | 0.30 / 10.9 | 0.38 / 8.71 | 0.49 / 2.45 |
| 9.3 | CO$^{+++}$ ? | – | – | 1.21 / 52.0 | 0.91 / 87.8 | 0.83 / 67.5 | 0.91 / 33.5 | 0.95 / 45.8 |
| 12 | C$^+$ | – | – | 0.26 / 1.28 | 0.56 / 0.50 | 0.49 / 0.40 | 0.93 / 0.98 | 1.19 / 0.98 |
| 13 | CH$^+$ | – | – | – | 0.30 / 1.18 | 0.30 / 1.04 | 0.49 / 1.04 | 0.47 / 0.16 |
| 14 | CH$_2^+$; CO$^{++}$ | 26.7 / 0.07 | 24.6 / 0.11 | 22.6 / 0.72 | 20.6 / 0.72 | 17.7 / 0.61 | 17.1 / 0.43 | 18.2 / 0.96 |
| 16 | O$^+$ | 20.9 / 0.30 | 20.0 / 1.99 | 16.8 / 1.99 | 15.9 / 1.65 | 13.1 / 0.96 | 13.4 / 1.22 | 14.9 / 1.50 |
| 17 | OH$^+$ | – | – | – | 0.45 / 0.06 | 0.49 / 0.17 | 0.59 / 0.13 | 0.75 / 0.17 |
| 18 | H$_2$O$^+$ | – | 0.66 / 0.12 | 0.25 / 0.05 | 0.55 / 0.05 | 0.67 / 0.03 | 0.64 / 0.03 | 0.60 / 0.05 |
| 28 | CO$^+$ | 37.9 / 0.02 | 37.4 / 0.02 | 39.9 / 0.02 | 37.7 / 0.02 | 41.1 / 0.02 | 39.2 / 0.02 | 37.3 / 0.03 |
| 29 | HCO$^+$ | – | 0.64 / 0.10 | 0.31 / 0.03 | 2.89 / 0.13 | 3.08 / 0.13 | 3.24 / 0.17 | 3.19 / 0.16 |
| 32 | O$_2^+$ | 14.4 / 0.03 | 13.4 / 0.03 | 14.1 / 0.03 | 13.2 / 0.02 | 15.5 / 0.02 | 14.4 / 0.02 | 13.3 / 0.01 |
| 44 | CO$_2^+$ | – | – | 0.16 / 0.03 | 0.63 / 0.02 | 0.56 / 0.02 | 0.72 / 0.02 | 0.75 / 0.03 |
| 45 | COOH$^+$ | – | 0.76 / 0.01 | 0.14 / 0.16 | 1.18 / 0.03 | 1.26 / 0.03 | 1.29 / 0.02 | 1.19 / 0.03 |
| 46 | HCOOH$^+$ | – | 0.14 / 0.11 | 0.11 / 0.06 | 0.91 / 0.01 | 0.97 / 0.01 | 1.02 / 0.02 | 0.94 / 0.01 |

Table 1. Relative intensities (partial ion yield – PIY) and kinetic energy $U_0$ release by fragments in the formic acid mass spectra, as a function of photon energy. Only fragments with intensity >0.1% were tabulated. The estimated experimental error was 10% (from Boechat-Roberty et al. 2005)

In order to test the above equation, we measured the argon mass spectrum under the same conditions. The kinetic energy value achieved for the Ar+ ions is in agreement with the mean kinetic energy $(3/2)KT$ obtained assuming Maxwell's distribution law. For example, the calculated values for kinetic energy release ($U_0$) of formic acid fragments produced by the interaction with soft X-ray photons in the range from 200 eV to 310 eV is shown in Table 1 (from Boechat-Roberty et al. 2005). We observe that, in the case of formic acid, the highest kinetic energy release was associated with the lightest fragment H+ ($m/q$ = 1), as expected. Extremely fast ionic fragments ($U_0$ > 10 eV), usually associated with dissociation of doubly or multiply-charged ions, were also observed at high photon energies. These observations point to the important role of Auger process ion fragmentation of core-ionized polyatomic molecules. The Coulomb explosion associated with the Auger process should explain the increase in kinetic energy of the ionic fragments, reflected in the increasing broadening of several fragments. This broadening observed in simple coincidence spectra (PEPICO) and its consequence on the shape of peaks in mass spectra was discussed by Simon et al. (1993).

### 2.1.2 Multicoincidence spectra (PE2PICO, PE3PICO)

In addition to the PEPICO spectra, two kinds of other coincidence mass spectra can be obtained simultaneously, PE2PICO spectra (PhotoElectron Photoion Photoion Coincidence) and PE3PICO spectra (PhotoElectron Photoion Photoion Photoion Coincidence). These spectra have ions from double and triple ionization processes, respectively, which arrive coincidentally with photoelectrons. Of all signals received by the detectors, only about 10% come from PE2PICO and 1% from PE3PICO spectra, indicating that the majority of the contribution is due to single event coincidence. Typical PE2PICO and PE3PICO spectra are given in Figure 16a and 16b, respectively. Inset figures show the *coincidence figures* (parallelogram or elliptical shape-like set of points) of a given set of fragments. The slopes of the coincidence figures in PE2PICO and PE3PICO spectra provide the information about the dissociation mechanism of the fragments that reach the detector in coincidence with the photoelectron (Pilling et al. 2007b; Pilling et al 2007c).

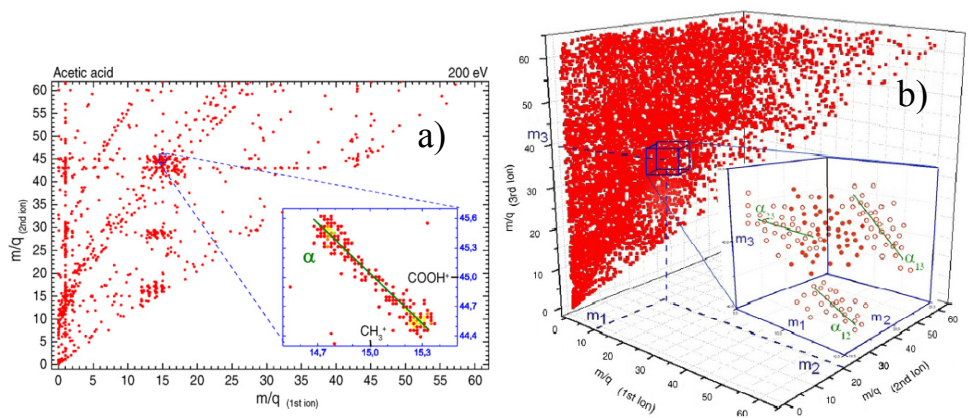

Fig. 16. a) Typical PE2PICO spectra. b) Typical PE3PICO spectra. The slopes of the coincidence figures in PE2PICO and PE3PICO spectra give the information about the dissociation mechanism of selected fragments.

The multi-coincidence detection technique was introduced by Fransiski et al. (1986) and Eland et al. (1986, 1987) in experiments involving energies at valence level near the ionization of inner shell electrons. In such regions, the probability of multiple ionizations increases due to the ionization processes such as Auger. In PE2PICO, the photoexitation/photoionizaton mechanism results a double ionization of parent molecule ($M^{++}$) which can dissociate or not. A typical reaction is illustrated by

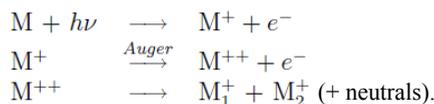

$$M + h\nu \longrightarrow M^+ + e^-$$
$$M^+ \xrightarrow{Auger} M^{++} + e^-$$
$$M^{++} \longrightarrow M_1^+ + M_2^+ \text{ (+ neutrals)}.$$

The production of negative ions as the result of multi-ionized species is also observed but this probability is negligible compared with positively charged ions.

Following Eland (1989) and Simon et al. (1993) the dissociation dynamics of double charge species can be studied by analyzing the coincidence figure (see inset Fig. 16a) of the ions in double coincidence mode in PE2PICO spectra. These authors suggested several mechanisms that may occur from the ionization with soft X-rays photons. Some examples are given below:

i. Two Body dissociation: This is the simplest dissociation case: The slope of coincidence figure (see inset figure 16a) is exactly - 45 degrees ($\alpha=-1$).

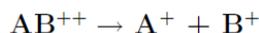

$$AB^{++} \rightarrow A^+ + B^+$$

ii. Tree body dissociation: In this situation a third neutral fragment is produced from the photodissociation. Some examples are given below:
- Concerted dissociation: In this case, the kinetic energy distribution (or momentum) between the three bodies in not necessarily unique. This gives one coincidence figure with a ovoid (or circular) shape, in particular if the momentum of the fragments were not aligned (Eland et al. 1986).

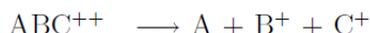

$$ABC^{++} \longrightarrow A + B^+ + C^+$$

- Deferred charge separation: In this case, the dissociation occurs in two steps. In the first the double charge molecule is broke into two species, but the charge is retained in one of the instable fragment. After that, this double charged fragment dissociates into two small single charge fragments.

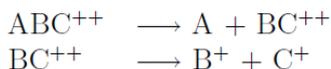

$$ABC^{++} \longrightarrow A + BC^{++}$$
$$BC^{++} \longrightarrow B^+ + C^+$$

The kinetic energy released in the last reaction (columbic explosion) is much higher than involved in the first step. Therefore, the momenta of the fragments can be treated as $p_A \approx 0$, and $p_B = -p_C$. As in the two body dissociation, this results in a coincidence figure with exactly - 45 degrees ($\alpha=-1$).
- Secondary decay: In this situation the molecular ion first brakes in two charged fragments and then one of these fragments brakes producing a neutral fragment and another small charged species.

$$\begin{aligned} ABC^{++} &\longrightarrow B^+ + AC^+ \\ AC^+ &\longrightarrow A + C^+ \end{aligned}$$

Following Simon et al. (1993), the slope of coincidence figure in PE2PICO spectra in this case is given by:

$$\alpha = -\frac{m_c}{m_a + m_c} \text{ for } m_b > m_c \quad \text{or} \quad \alpha = -\frac{m_a + m_c}{m_c} \text{ for } m_b < m_c$$

In PE3PICO, the photoexitation/photoionizaton mechanism results a triple ionization of parent molecule ($M^{+++}$) which can dissociate or not. A typical reaction is illustrated by

$$\begin{aligned} M + h\nu &\longrightarrow M^+ + e^- \\ M^+ &\xrightarrow{Auger} M^{+++} + 2e^- \\ M^{+++} &\longrightarrow M_1^+ + M_2^+ + M_3^+ \text{ (+ neutrals)}. \end{aligned}$$

In this case, the coincidence figure involving a given set of three fragments has three dimensions (See inset Fig. 16b). Some information about the dissociation mechanism involving these fragments can be given by analyzing the projection of its coincidence figure projected in each plane, however, the dissociation mechanism in this case is difficult to determine.

From the analysis of the fragment's produced mass spectra, we determined the branching ratio of the fragments. Employing a similar methodology, as previously discussed, the photodissociation and photoionization cross sections in double and triple ionization can also be determined.

The relative intensities of each ion pair coincidences in each spectrum can be determined directly from spectra analysis. In the case of PE2PICO spectra, we obtained the partial double coincidence yield (PDCY) by:

$$\text{PDCY}_{i,j} = \left( \frac{A_{i,j}}{A_t^{2+}} \pm \frac{\sqrt{A_{i,j}} + A_{i,j} \times \text{ER}/100}{A_t^{2+}} \right) \times 100\% \quad (13)$$

where $A_{i,j}$ is the number of events in double coincidence of a given $i$ and $j$ ion pair, the $A_t^{2+}$ is the total number of counts of PE2PICO spectra, and ER = 2–4% is the estimated uncertainty due to the data acquisition and the data treatment (Pilling et al 2007b).

In a similar manner, for the PE3PICO spectra, we determined the Partial Triple Coincidence Yield (PTCY) (Pilling et al 2007b, Pilling et al 2007c).

From the cross section, we can derive the photo-dissociation rate and photo-production rate (of a given fragment), if we know the soft X-ray flux in a give astrophysical region. For example, the $H_3^+$ photoproduction rate due to the dissociation of methyl compound molecules by soft X-rays (200–310 eV) is given by the simple expression

$$k_{\text{ph}} = \int \sigma_{H_3^+}(\varepsilon) F(\varepsilon) \, d\varepsilon \sim \sigma_{H_3^+} F_{\text{softX}} \quad (s^{-1}) \quad (15)$$

where $\sigma_{H_3^+} = \sigma_{H_3^+}^+ + \sigma_{H_3^+}^{++}$ and $F_{softX}$ is the averaged H$^{+3}$ photoproduction cross-section and photon flux over the soft X-ray energy (200–310 eV), respectively (Pilling et al 2007d).

The photodissociation rate can be obtained employing the photodissociation cross section $\sigma_{ph-d}$ and soft X-ray photon flux in the given astrophysical scenario.

$$k_{ph-d} = \int_{softX} \sigma_{ph-d}(\varepsilon) F(\varepsilon) d\varepsilon \approx \sigma_{ph-d} \times F_{softX} \ (s) \qquad (16)$$

The half-life ($t_{1/2}$) of the studied compounds in the presence of interstellar soft X-rays can be obtained directly from

$$t_{1/2} = \frac{\ln 2}{k_{ph-d}} \approx \frac{0.69}{\sigma_{ph-d} \times F_{softX}} \ (s) \qquad (17)$$

**2.2 Condensed-phase (ice) and solid-phase experiments**

In the cold and dense interstellar regions in which stars are formed, CO, $CO_2$, $H_2O$, and other molecules collide with and stick to cold (sub)micron-sized silicate/carbon particles, resulting in icy mantles. Inside these regions, called molecular clouds, the dominant energy source is cosmic rays and soft X-rays from embedded sources or the interstellar radiation field. These kinds of radiation can also produce an expressive amount of fast electrons, which can also promote chemical differentiation on ices in a different way when compared with photons.

Several experiments employing the interaction between soft X-rays and frozen (or solid) compounds were performed to investigate the effects of stellar ionizing radiation in astrophysical ices. Two different techniques were employed, one to investigate the species, which desorbs from the surface due to the irradiation, and another to analyze the bulk (of the ice) during the irradiation.

**2.2.1 Photo stimulation ion desorption**

In an attempt to investigate the fragments that were released into gas phase due to the impact of soft X-rays on astrophysical ices, we employed the photo stimulation ion desorption (PSID) technique by using the spherical grating monochromator (SGM) beam line in single-bunch mode (pulse period of 311 ns, with a width of 60 ps) at the Brazilian synchrotron light source (LNLS). By using soft X-rays photons, one can excite specific atoms inside a molecule by tuning the incoming radiation as a consequence of the different chemical shifts. Therefore, this technique is element- and site-specific. Several species were investigated by this technique including formic acid and methanol.

Formic acid (HCOOH) is the simplest carboxylic acid, while methanol ($CH_3OH$) is the simplest alcohol. These molecules have been found abundantly in icy mantles on interstellar and protostellar dust grains. These small molecules together with glycine (the smallest amino acid) serve as model systems to understand the properties of larger and more complex amino acids and proteins. Formic acid, for example, has been observed in several astronomical sources such as comets (Kuan et al. 2003, Snyder et al. 2005, Jones et al. 2007), protostellar ices NGC 7538:IRS9 (Cunningham et al. 2007), condritic meteorites (Cronin and Pizzarrelo, 1997), dark molecular clouds (Turner, 1991), and regions associated with stellar formation (Nummelin et al. 2000; Ehrenfreund et al. 2001).

In some massive star-forming regions such as Sgr B2, Orion KL, and W51, formic acid has been observed. Kuan et al (2003) suggest that glycine may be detected in these regions, but more observation are necessary to confirm this possibility Methanol has been detected through infrared spectroscopy in some low- and high-mass protostars such as W33A and RAFGL 7009. In these regions, $CH_3OH$ is the most abundant solid-state molecule after $H_2O$ (Dartois et al. 1999; Pontoppidan et al. 2003; Boogert et al. 2008).

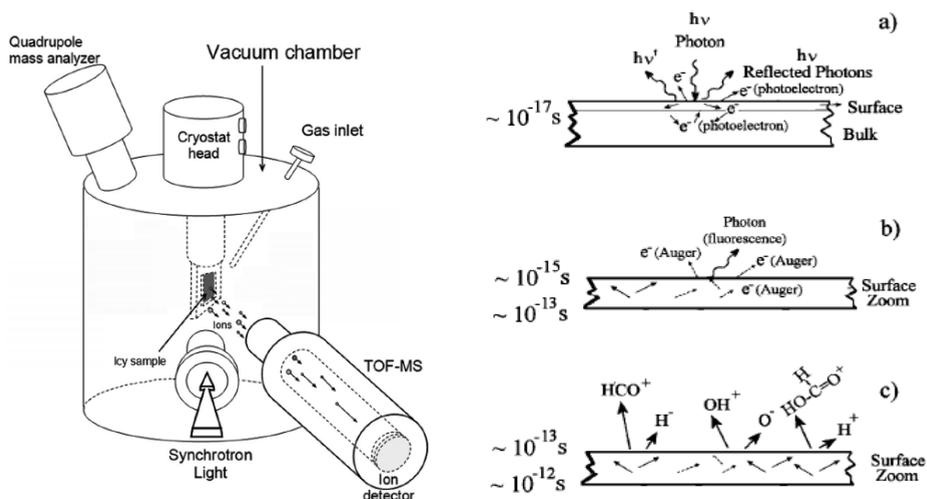

Fig. 17. Schematic diagram of the experimental set-up that was employed in the photo stimulation ion desorption of astrophysical ice analogs (left). Scheme of desorption process due to X-ray impact on the ice surface, simulating the interstellar ice (right).

When X-ray photons interact with the grain surfaces, the icy molecules can dissociate, producing small neutral or ionized species and atoms. If the surface temperatures are around 50 K, the radicals can diffuse and associate to form larger molecules. Afterwards, these neutral and ionic species can sublimate from the surface by thermal or non-thermal desorption mechanisms. The non-thermal desorption mechanisms are those stimulated by photons or energetic charged particles. Each ionizing agent will promote a different fragmentation in the molecule, favoring the formation of ionic species rather than neutral species, depending on the impact energies. The inner-shell photoexcitation process, for example, may produce instabilities on the molecular structure leading to peculiar dissociation pathways. The molecules and ions formed and desorbed into the gas phase, as well as the complex molecules formed from these species, will depend on the interaction of radiation field in the considered region. To quantify the complex organic molecules incorporated in grains or desorbed into the gas phase and to predict the chemical evolution, it is necessary to establish the main formation pathways, which can be tested in laboratories. Therefore, studying the effects of different ionization agents on the ices is crucial in the knowledge of evolution of interstellar chemistry.

In this chapter, the main dissociate channel in the molecules are produced by hole excitation/ionization. When grains are processed by soft X-ray photons, the photons can be transmitted, absorbed, reflected, or scattered by the surface. Figure 18 shows a schematic diagram of desorption process by X-ray photons on the ice surface, simulating the effects of the radiation on interstellar/cometary ices. In Figure (17a), an X-ray photon is absorbed ($t \approx 10^{-17}$ s), promoting an electron into an excited state or to the continuum, creating a core hole (a hole in an internal orbital of the atom) on the molecule. As shown in Figure 17b, after relaxation, X-ray fluorescence or Auger electron emission can occur, depending on the atomic number of the atom involved. If the atomic number of the involved atom is below ~16, the Auger decay has probability ~1. On the other hand, if the atomic number is above 16 the X-ray fluorescence process can occur with high probability (Jenkins 1999). For molecules such as formic acid and methanol, which will be shown here, the atoms have a low atomic number. In this case, the probability of fluorescence vanishes and the Auger decay is dominant (Andrade et al, 2010). Mainly because of Coulomb repulsion of positive holes in valence orbitals, molecular dissociation and desorption occur (Figure (17c). The soft X-ray photons interact directly with surface molecules. In this way, the fragments are desorbed mainly from the interaction region. The energy deposited into the system is enough to produce resonant core level transition, which means localized excitation. The final step in this process is bond scission at the site of excitation and consequently desorption.

The experimental setup, presented here, includes a sample manipulator (where the molecules are condensed onto an Au thin film) connected at a helium cryostat and a time-of-flight mass spectrometer (TOF-MS) housed in an Ultra High Vacuum (UHV) chamber with a base pressure around $10^{-9}$ Torr, which increased by two orders of magnitude (1.5x $10^{-7}$ mbar) when the gas was introduced for condensation for around six min. The TOF-MS consists basically of an electrostatic ion extraction system, a drift tube and a pair of microchannel plate (MCP) detectors, disposed in the Chevron configuration. After extraction, the ions (positive and negative) travel through three metallic grids before reaching the MCP. The grids have a nominal transmission of ~90 per cent.

The time-of-flight (TOF) of a specific desorbed ion is given by its position in the spectrum plus an integer multiple of the synchrotron radiation (SR) pulse interval (311 ns) corresponding to the number of cycles that have passed from desorption of the ion to its detection. The mass resolution was limited by the resolution of the TOF spectrometer and by the time resolution of the TDC (250 ps ch$^{-1}$). The typical measured full width at half-maximum FWHM found for ionic species was 1.5 < FWHM < 20 ns, being 1.5 for H$^+$. In the photon case, the FWHM was ~250 ps.

The attribution of PSID spectra obtained in single-bunch mode is not an easy task. The time window of 311 ns is small enough even for the hydrogen ion. Thus H$^+$ and heavier fragments are detected at different cycles (or starts) of the experiment, causing the overlapping of many different species. However, because of the very good reproducibility of the SR pulses and the good resolution achieved in these experiments, it was possible to identify different structures and suggest the assignment of the TOF spectra.

In order to determine the photodesorption yield (Y) for the ions from condensed molecules, we obtained the area of each peak. $Y_i$ is the number of ions desorbed per incident photon for each ion $i$ (equation 18) and it was determined by dividing the peak area ($A$) by the number of bunches ($N_b$) and by the number of photons per bunch ($n_{ph}$):

$$Y_i = \frac{A}{N_b \, n_{\text{ph}}}. \tag{18}$$

The number of photons incident on a surface of 2 mm² during 60 ps (duration of the pulse), at the single-bunch mode, is about 1550 photons bunch⁻¹, which corresponds to a photon flux of $5 \times 10^{11}$ photons s⁻¹ cm⁻². The errors associated with ion yield measurements depend on peak position. For single peaks, the error was around 10 per cent. For blended peaks, the estimated errors were as high as 35 %.

The icy sample temperature was around 55-56 K during all measurements. The ice thickness was in the micrometer range. Due to the temperature and pressure conditions in the chamber, the formed ice was expected to be in its amorphous phase, similar to the icy organic found in comets and in the interstellar medium.

Figure 19 shows the total ion yield (TIY) of Near-edge X-ray Absorption Fine Structure (NEXAFS) spectrum at the oxygen 1s-edge for condensed formic acid (HCOOH), covering the photon energy range from 528 to 553 eV (Andrade et al. 2009). The NEXAFS spectrum has high-pitched peaks (B and D) and one broad band (F), which are related to electronic transitions from the oxygen 1s electron to unoccupied molecular orbitals. In comparison with gas phase NEXAFS data on formic acid (Prince at al. 2002; Hergenhahn et al. 2003), the peaks B can be assigned in the figure. The broad band F is located above the ionization threshold.

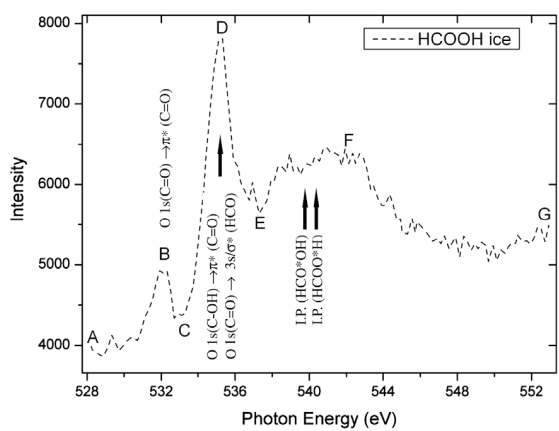

Fig. 19. The total ion yield (TIY) of Near-edge X-ray Absorption Fine Structure (NEXAFS) spectrum at the oxygen 1s-edge for condensed formic acid, covering the photon energy range from 528 to 553 eV.

Figure 20 shows typical positive (a) and negative (b) time-of-flight mass spectra of condensed formic acid obtained at the excitation energy labeled D (535.1 eV) (Andrade et al. 2009) and (c) positive ions and (d) negative ions TOF spectra from methanol ice obtained at 537 eV (Andrade et al. 2010). In both cases, $CH_2^+$ and/or $CO_2^+$ represent the main ionic molecular signal desorbed after photon excitation at the O 1s-edge followed by intense H⁺. In the negative spectra, only H⁻ and O⁻ appear. Heavier fragments with lower intensities, such as C⁺, CH⁺, O⁺, CO⁺, and HCO⁺ were found in the HCOOH spectra, and the (HCOOH)H⁺ cluster is probably present. In the methanol case, $CH_3^+$ species was also formed with low intensity, and unlike in the case of formic acid, CH⁺ was not seen.

Measurements of TOF spectra from HCOOH at different photon energies (A–G) around the O 1s-edge were also performed in an attempt to understand the ionic desorption process from condensed formic acid. Table 2 shows the edge-jump (the $I_{above}/I_{below}$ ratio) values for the total ion yield photoabsorption spectrum for several ions, where $I_{above}$ and $I_{below}$ represent the intensity signals above and below the absorption edge, respectively. Moreover, this table shows the positive and negative desorption yield (desorbed ion for photon incident) to the formic acid and methanol.

In the HCOOH as well as the $CH_3OH$ spectra, $OH^+$ ion peak overlaps with $H_3^+$ and $C_2^+$ peaks, thus it is difficult to separate their contributions. In the case of HCOOH, these species, as well as the $CH_2^+$ (and/or $CO_2^+$), $CO^+$, and $C_2^+$ do not show a significant increase at the D resonance, suggesting that the X-ray induced electron stimulated desorption (XESD) process, which competes with the ASID (Auger stimulated ion desorption) process, is responsible for the desorption signal. In the XESD process, desorption of surface species occurs due to outgoing energetic Auger and photoelectrons, most of which originate in the bulk (Ramaker et al. 1988; Purdie et al. 1991). The XESD process is proportional to the total ion yield and the desorbed ions appear below and above the absorption edge. Its ratios (Table 2) follow those of the photoabsorption spectrum.

Taking into account the positive and negative desorbed ions from formic acid, some possibilities for the $O^-$ and $H^-$ formation from singly charged desorbed ions are suggested below (from the ions seen in the positive ion spectra) (Andrade et al. 2009):

$$HCOOH^+ \rightarrow O^- + HCO^+ + H^+ \quad (1)$$
$$\rightarrow H^- + C^+ + H^+ + O_2 \quad (2)$$
$$\rightarrow O^- + CH^+ + H^+ + O \quad (3)$$
$$\rightarrow H^- + CH^+ + O^+ + H \quad (4)$$
$$\rightarrow O^- + C^+ + H_2 + O^+ \quad (5)$$
$$\rightarrow H^- + HCO^+ + O^+ \quad (6)$$
$$\rightarrow O^- + C^+ + H^+ + O \quad (7)$$

It is believed that the suggested reactions contribute most to formic acid dissociation, because we are working around the O 1s-edge. Thus, mechanisms like electron capture are not considered in our discussion and non-detected positive ions do not participate in the above reactions. OH species is probably released in the neutral form or HCOOH breaks giving H and O ions, rather than OH cations and anions, because this species showed low intensity in the formic acid positive spectra and was absence in the negative spectra. Because $O^+$, $O^-$, and $H^-$ yields are very intense at D resonance, this second option appears favorable. In the gaseous phase, the formation of the anionic OH fragment, via resonant excitation at the carbon 1s-edge and its complete absence near the oxygen 1s-edge was reported by (Stolte et al. 2002). $C_2^+$ was detected with a weak intensity in the formic acid spectra and large amounts of $CH_2^+$, $HCO^+$, $H^+$, and $O^+$ were present, but all of these fragments can be formed due to many different pathways, not only as counterparts of $O^-$ and $H^-$.

By comparing Table 2 and the cation yields for $O^-$ and $H^-$, we observe that $O^+$, $C^+$, and $CH^+$ have similar behavior as these anions at the D resonance, suggesting that equations (1)–(7) are strong candidates for $H^-$ and $O^-$ formation at O 1s-edge excitation. In addition, taking into account that $C^+$, $CH^+$, and $O^+$ ions show enrichment at the D resonance, the $O^-$ and $H^-$ formation from the reactions (4) and (5) seems to be more favorable (Andrade et al. 2009).

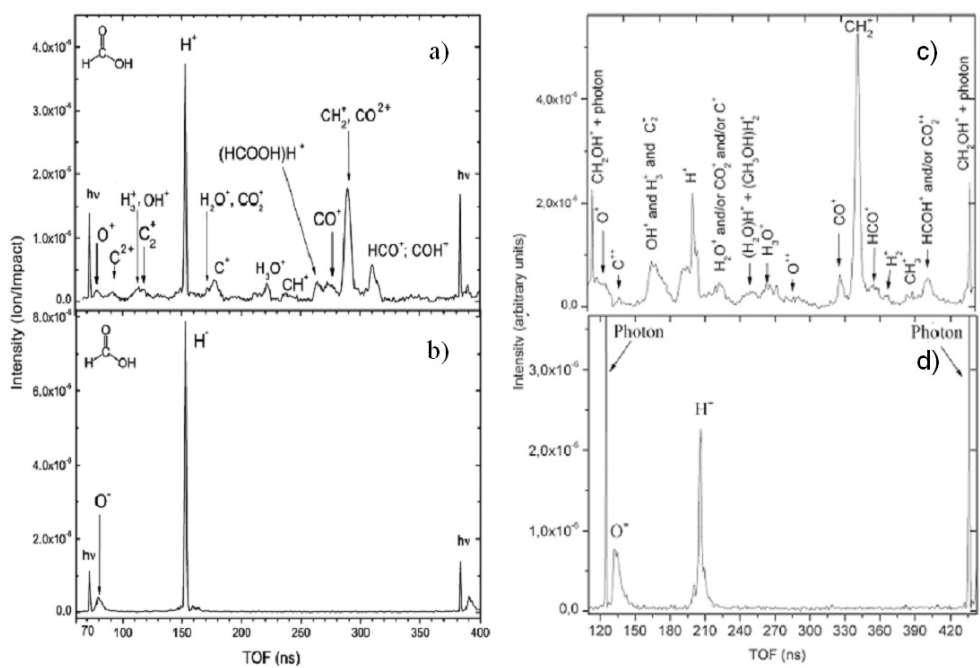

Fig. 20. a) positive ions and b) negative ions TOF spectra of condensed HCOOH obtained at 535.1 eV energy and c) positive ions and d) negative ions TOF spectra from methanol ice obtained at 537 eV.

| Ion assignment | ID/IB (HCOOH) | ID/IF (HCOOH) | $Y_i$ (x10$^{-9}$) HCOOH | $Y_i$ (x10$^{-9}$) $CH_3OH$ |
|---|---|---|---|---|
| NEXAFS (TIY) | 1.60 | 1.22 | | |
| $H^+$ | 1.31 | 0.76 | 2.88 | 10.26 |
| $O^+$ | 2.33 | 1.88 | 1.14 | 5.006 |
| $C_2^+/OH^+/H_3^+$ | 1.82 | 1.22 | 8.45 | - |
| $C^+$ | 2.20 | 1.79 | 1.83 | - |
| $CH^+$ | 1.90 | 1.95 | 0.65 | - |
| $CH_2^+/CO^{2+}$ | 1.36 | 1.29 | 11.5 | 19.10 |
| $(HCOOH)H^+$ | 1.21 | 0.72 | 0.82 | - |
| $HCO^+$ | 1.47 | 0.91 | 1.39 | 1.619 |
| $CO^+$ | 1.47 | 1.11 | 0.47 | 1.342 |
| $H^-$ | 3.28 | 1.51 | 11.0 | 5.550 |
| $O^-$ | 2.25 | 1.22 | 1.70 | 4.832 |

Table 2. Edge-jump (the $I_{above}/I_{below}$ ratio) values for the total ion yield photoabsorption spectrum for several ions and $Y_i$ desorption ion yield (desorbed ion for photon incident) to HCOOH and $CH_3OH$ ice.

After photon impact, desorption from the surface of multiple charged ions is not the most probable event because fast neutralization compete efficiently with ion desorption. In addition, there is the possibility of reactions involving doubly charged ions. $S_2^+$ was measured in the photofragmentation study of poly(3-methylthiophene) (Rocco et al. 2004; Rocco et al. 2006) and condensed thiophene (Rocco et al. 2007) following sulphur K-shell excitation and $Cl_2^+$ was observed from solid $CCl_4$ at the Cl 1s → σ*(7a$_1$) excitation (Babba et al. 1997). To explain the primary H- and O- production mechanism from doubly-charged ions, the most produced positive ion yields are that of $CO^{2+}$ and $C^{2+}$, which are possibly present in the positive PSID spectrum (reactions (8)–(10)). However, enrichment of neither $CO^{2+}$ nor $C^{2+}$ associated with H- and O- increase at the D resonance is seen, suggesting that H- and O- formation from formic acid occur mainly from (4) and (5). In the case of methanol, $CO^{2+}$ and $O^{2+}$ are detected and $C^{2+}$ cannot be excluded because its peak overlaps with $CH_2^+$, which is predicted to be a more abundant product.

$$HCOOH^+ \rightarrow H + CO^{2+} + OH \quad (8)$$
$$\rightarrow O^- + C^{2+} + OH + H \quad (9)$$
$$\rightarrow H^- + C^{2+} + O_2 + H \quad (10)$$

Neither $COOH^{\pm}$ nor $HCOOH^{\pm}$ is formed from HCOOH fragmentation. Therefore, if an O 1s electron is excited, HCOOH dissociate into smaller or atomic fragments. Pelc et al. (2002) showed, from theoretical calculations, that it is not possible by the thermodynamical point of view to bind an extra electron to HCOOH, because it has a negative adiabatic electron affinity. In contrast to comparable gas-phase studies, $CH_3OH^+$ was not detected in the methanol positive mass spectra. This implies that either $CH_3OH^+$ is less stable in the ice phase compared to in the gas phase or that $CH_3OH^+$ is produced in the ice but lacks sufficient energy to desorb and thus escapes detection. Ionized formaldehyde, $H_2CO^+$, could also be identified and is intense in our methanol positive ion spectrum. This species has also been identified by other authors, in neutral form, using Fourier transform infrared spectroscopy (FTIR; Gerakines et al. 1996; Hudson & Moore 2000; Bennett et al. 2007). Palumbo et al. (1999) also observed a band at 1720 cm$^{-1}$ after ion irradiation with 3 keV helium ions in pure methanol, but admitted that there could be other species, such as acetone, contributing to this feature. Acetone was not found in our spectra.

$$CH_3OH^+ \rightarrow CH_2^+ + H^- + OH^+ \quad (11)$$
$$\rightarrow O^- + H^+ + CH_2^+ + H \quad (12)$$
$$\rightarrow H^- + HCOH^+ + H^+ \quad (13)$$
$$\rightarrow H^+ + H_2 + H^- + CO^+ \quad (14)$$
$$\rightarrow H_3^+ + H^- + CO^+ \quad (15)$$
$$\rightarrow HCO^+ + H^- + H_2^+ \quad (16)$$
$$\rightarrow C^+ + OH + H^- + H^+ \quad (17)$$
$$\rightarrow C^+ + O^- + H^+ + H_2 \quad (18)$$
$$\rightarrow C^+ + O^- + H + H_2^+ \quad (19)$$
$$\rightarrow C^+ + O^- + H + H_2^+ \quad (20)$$
$$\rightarrow O^+ + H^- + CH_3^+ \quad (21)$$
$$\rightarrow O^- + H^+ + CH_3^+ \quad (22)$$

Table 3. Possible pathways for H− and O− formation from singly charged desorbed ions from methanol ice.

Some possible pathways for the O$^-$ and H$^-$ formation from singly charged desorbed ions, which were observed in the methanol positive ion spectra, are suggested in Table 3. For the reaction pathways, only the detected positive ions were taken into account and other mechanisms, such as electron capture, were not considered in our discussion. Examining the cation yields in Table 2, we suggest that the equations (11)–(13) are strong candidates for H$^-$ and O$^-$ formation from methanol at the O1s excitation energy. Moreover, taking into account that C$^+$ and CH$_3^+$ ions show low abundance, the O$^-$ and H$^-$ formation from the (18)–(22) reactions seems to be less favorable.

In the cases of HCOOH and CH$_3$OH, the fragmentation pattern of the positive ions is clearly more pronounced than that of the negative ions, showing that positive ions are more easily formed than the negative one.

The astrophysical implication of the x-ray photon flux and the production of ions in the protoplanetary discs will be discussed. A protoplanetary disc is a rotating circumstellar disk of matter, including dust and gas, that surrounds a very young star and from which planets may eventually form or be in the process of forming, representing the early stages of planetary system formation. Measurements of high and intense X-ray fluxes from young stars have been obtained by Space telescopes such as *ROSAT*, *Chandra*, and *XMM–Newton*. These fluxes can be up to three orders of magnitude higher than for main-sequence stars (Gorti et al. 2009).

The closest known young T Tauri star is TW Hya, which has a massive, face-on optically thick disc. Using the infrared spectrograph onboard the *Spitzer* space telescope, Najita et al. (2010) suggested that a planet was probably formed in the inner region of the disc, at approximately 5 au. and also showed that TW Hya has a rich spectrum of emission of atomic ions (Ne II and Ne III) and molecules resulting from K-shell (Ne 1s) photoionization of neutral neon by X-rays.

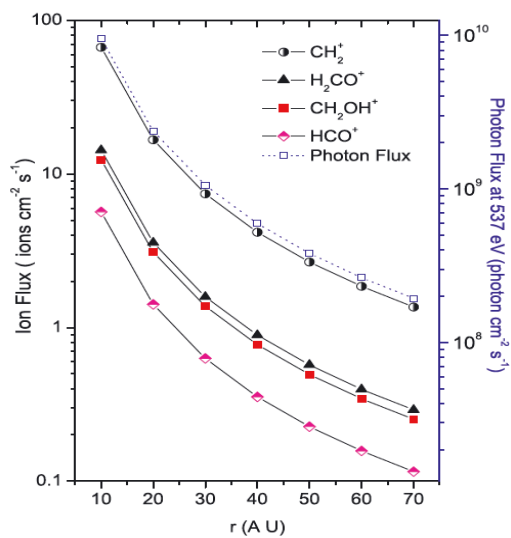

Fig. 21. Ion flux desorbed from icy methanol and X-ray at 537 eV photon flux in TW Hya as a function of the distance from a point in the disc to central star (from equation 19 for $\tau = 1$) (Andrade et al. 2010).

TW Hya has a X-ray luminosity ($L_X$) of $2 \times 10^{30}$ erg s$^{-1}$, integrated from 0.2 to 2 KeV (Kastner et al. 2002). At a specific energy the photon flux $F_X$ can be obtained by:

$$F_X = \frac{L_X}{h_\nu \cdot 4\pi r^2} e^{-\tau} \qquad (19)$$

where $\tau$ is the X-ray optical depth given by $\tau = \sigma_{ab}(E) N_H$ and $r = (R^2 + z^2)^{0.5}$ is the distance from a point in the disc to the star in centimeter. $R$ is the distance from the star to disc and $z$ is the height from the mid-plane (Gorti & Hollenback 2004). $\sigma_{ab}(E)$ is the absorption cross-section as a function of the photon energy $E$, and $N_H$ is the hydrogen column density obtained by the product of the number density ($nH$) and the distance from the central star (Andrade et al. 2010). It was assumed that $\tau = 1$ using the $\sigma_{ab}(E)$ per H atom given by Gorti & Hollenback (2004) model and $N_H \sim 10^{21}$ cm$^{-2}$. $nH$ varies from $6.6 \times 10^6$ to $0.95 \times 10^6$ cm$^{-3}$ in the distances range from 10 to 70 au. The ion flux (ions cm$^{-2}$s$^{-1}$) can be estimated by $f_{ion} = F_X Y_i$ ($Y_i$ is the photodesorption yield, ions/photon) from equation (18). Figure 21 shows the photon flux at 537 eV in TW Hya and the flux of the most important ions which are desorbed from the methanol ice as a function of the distance from the central star.

The most abundant ion desorbed by methanol is the $CH_2^+$ followed by ionized formaldehyde and $CH_2OH^+$. For example, at 30 au, a flux of $\sim 10^9$ photons cm$^{-2}$ s$^{-1}$ may produce $\sim 9$ $CH_2^+$ ions cm$^{-2}$ s$^{-1}$ and desorb $\sim 0.6$ $HCO^+$ ion cm$^{-2}$ s$^{-1}$. Taking into account the half-life of the region ($1 \times 10^6$ yr), the amount of desorbed $HCO^+$ (formyl ion) and $CH_2^+$ will be around $2 \times 10^{14}$, and $3 \times 10^{15}$ ions cm$^{-2}$, respectively.

The ion production rate $R_{ion}$ (cm$^{-3}$ s$^{-1}$) from grain surface (eq. 20), is dependent on the ion flux $f_{ion}$, the amount of grains $n_{gr}$ (cm$^{-3}$), their surface cross section $\sigma_{gr}$ (in cm$^2$ which is equal to $\pi a^2$, where $a$ is the radius of the dust particles), and the fraction $X$ of the surface covered by methanol (which can be archived at a range from zero to unity) by the expression:

$$R_{ion} = n_{gr} \sigma_{gr} f_{ion} X \qquad (20)$$

Assuming that $n_{gr} \sim 10^4$ cm$^{-3}$ (Gorti et al. 2009), $a = 0.1$ µm, and $X = 1$ (the surface is totally covered by methanol), the $R_{HCO+}$ will be approximately $1 \times 10^{-6}$ cm$^{-3}$ s$^{-1}$. If $X = 0.3$ (70 per cent of the surface is covered by water ice), the $R_{HCO+}$ will be approximately $3 \times 10^{-7}$ cm$^{-3}$ s$^{-1}$. The half-life of the methanol can be obtained by Eq. 17, which does not depend on the molecular number density. Considering that the photodissociation cross-section at 537 eV is of the same order of magnitude as the photoabsorption cross-section for the gaseous methanol ($\sim 1,5 \times 10^{-18}$ cm$^2$, Ishii & Hitchcook, 1988), the methanol molecules survives $\sim 15$ years at 30 au.

**2.2.2 Photolysis of astrophysical ices analogs**

To complete the investigation about the effects of soft X-rays on the surface of interstellar ices, we also study the chemical evolution inside the ices by employing Fourier transform infrared (FTIR) spectrometry. In these experiments, *in-situ* analysis of the bulk of the astrophysical ice analogs are performed by a FTIR spectrometer, coupled to the experimental chamber, at different X-ray photon doses.

In these experiments, a gaseous sample (mixture) was deposited on a cold substrate (NaCl or CaF$_2$) coupled to closed-cycle cryostat, inside a high vacuum chamber and exposed to soft X-rays. Most of the samples were investigated at 13 K. Pure samples and several mixtures were investigated including N$_2$+CH$_4$, CO$_2$+H$_2$O, and CO+H$_2$O.

Figure 22 shows a schematic diagram of the experimental setup employing soft X-rays in solid samples of astrophysical interest. In the irradiation, the substrate with the ice sample is in front of the photon beam. After each irradiation dose, the target is rotated by 90° for FTIR analysis.

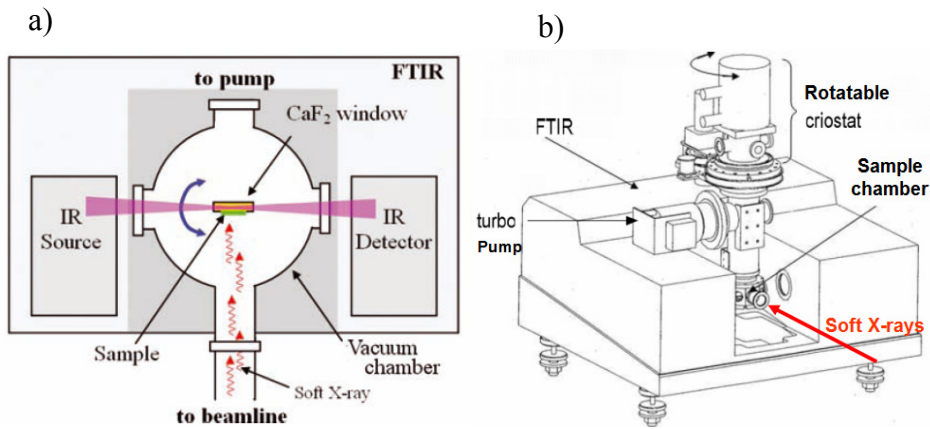

Fig. 22. Schematic diagram of the experimental setup employing soft X-rays on solid samples of astrophysical interest. a) top view. b) perspective view.

In the first set of experiments, we attempted to simulate the photochemistry induced by soft X-rays (from 0.1- 5keV) at aerosol analogs in the upper atmosphere of Titan, the largest moon of Saturn. The experiments were performed inside a high vacuum chamber coupled to the soft X-ray spectroscopy (SXS) beamline, at Brazilian Synchrotron Light Laboratory (LNLS), employing a continuum wavelength beam from visible to soft X-rays with a maximum flux in the 0.5-3 keV range. The frozen sample (Titan´s aerosol analog) was made by a frozen gas mixture containing 95% N$_2$ + 5% CH$_4$ (and traces of H$_2$O and CO$_2$). The in-situ analysis was done using FTIR spectrometry. Among the different chemicals species produced, we observe several nitrogen-rich compounds including one of the DNA nucleobases, the adenine molecule (Pilling et al. 2009). This detection was confirmed also by ex-situ chromatographic and RMN analysis of the organic residue produced. Unlike UV photons, the interaction of soft X-rays with interstellar ices produces secondary electrons in and on the surface, which allows different chemical pathways.

Figure 23a shows the evolution of simulated Titan aerosol at different exposure times up to 73h (1 h ~ 3 × 10$^{10}$ erg/cm$^2$). A comparison between FTIR spectra of 73 h irradiated sample at 15, 200, and 300 K is given in Figure 23b. The vertical dashed lines indicate the frequency of some vibration modes of crystalline adenine. Figure 23c shows a comparison between LNLS´s SXS beamline photon flux and the solar flux at the Titan orbit.

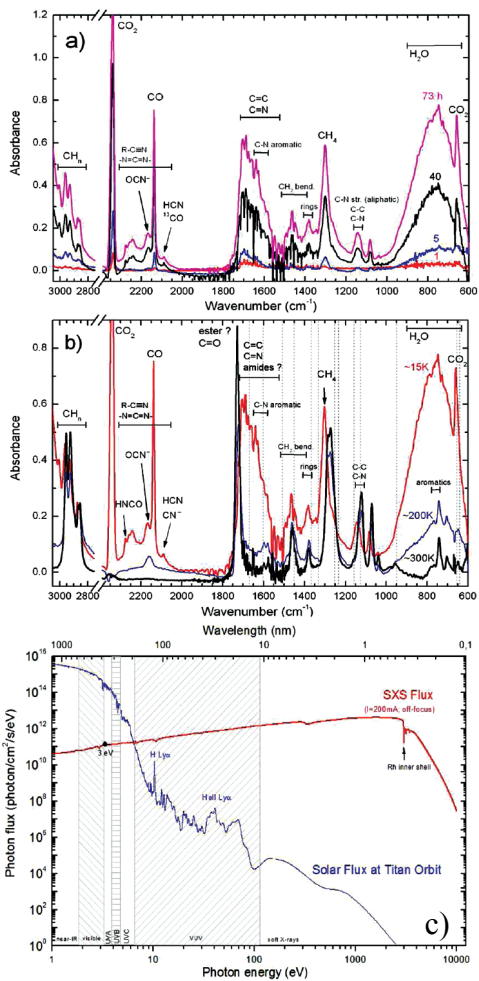

Fig. 23. a) FTIR spectra of the organic residue (tholin) produced by the irradiation of condensed Titan atmosphere analog at 15 K NaCl surface at different exposure times. b) Evolution of FTIR spectra of 73 h irradiated sample at three different temperatures: 15, 200, and 300 K (organic residue). c) Comparison between SXS beamline photon flux and the solar flux at the Titan orbit (adapted from Pilling et al 2009).

In a second experiment, we investigate the photochemistry effects of soft X-rays in CO and $CO_2$ ices covered by water ice (cap) within a soft X-ray field analogous to the ones founds in astrophysical environments. The measurements were taken at the Brazilian Synchrotron Light Laboratory at the soft X-ray spectroscopy (SXS) beamline, employing a continuum wavelength beam from visible to soft X-rays with a maximum flux between 0.5-3 keV range ($10^{12}$ photons cm$^{-2}$ s$^{-1}$ ~ $10^4$ ergs cm$^{-2}$ s$^{-1}$). Briefly, the samples were deposited onto a NaCl substrate cooled at 12 K under a high vacuum chamber ($10^{-7}$ mbar) and exposed to different radiation doses up to 3 h. In-situ sample analysis was performed by a Fourier transform infrared spectrometer (FTIR) coupled to the experimental chamber. Figure 24 shows a) the evolution of the abundance of selected species in the irradiation of CO + Water cap 12 K ice by soft X-rays; b) Evolution of the abundance of selected species in the irradiation of $CO_2$ + Water cap 12 K ice by soft X-rays and c) infrared spectra between 1600-600 cm-1 of the two samples after and before irradiation.

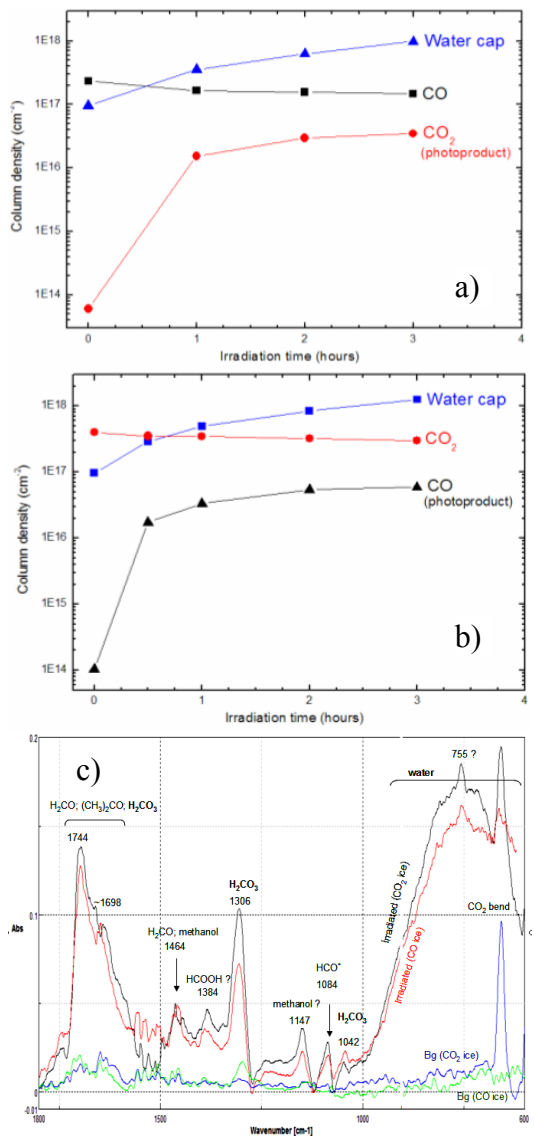

Fig. 24. a) Evolution of the abundance of selected species in the irradiation of CO + Water cap 12 K ice by soft X-rays. b) Evolution of the abundance of selected species in the irradiation of $CO_2$ + Water cap 12 K ice by soft X-rays. c) Infrared spectra between 1600-600 cm-1 of the two samples after and before irradiation.

The results showed that deep inside a typical dense cloud (e.g. AFGL 2591) the half-life of the studied species due to the photodissociation by penetrating X-rays photons was found to be about $10^3$ -$10^4$ years, which is in agreement with the chemical age of the cloud. Moreover, from the photodissociation of condensed CO, $CO_2$, and water molecules several organic species were also produced including $H_2CO$, $H_2CO_3$ and possibly formic acid and methanol. These results suggest that inside dense molecular clouds or other dense regions, where UV photons can not penetrate, the photochemistry promoted by soft X-rays photons (and cosmic rays) could be very active and plays an important role.

In these experiments, the analyzed samples were thin enough (i) to avoid saturation of the FTIR signal in transmission mode and (ii) to be fully crossed by the soft X-ray photon beam. The spectra were obtained in the 600 – 4000 $cm^{-1}$ wavenumber range with a resolution of 1 $cm^{-1}$. From the analysis of the IR spectra, we determined the amount of each molecular species present and its evolution as a function of soft X-ray dose. In addition, we determined the dissociation cross section of the sample and estimated the half-life of the studied species in solid-phase in astrophysical environments. For the low temperature experiments, we identified and quantified newly formed compounds.

In another experiment, we performed a photochemistry study of solid phase amino acids (glycine, DL-valine, DL-proline) and nucleobases (adenine and uracil) under a soft X-ray field in an attempt to test their stabilities against high ionizing photon field analogous to the ones found in dense molecular clouds and protostellar disks. Due to the large hydrogen column density and the amount of dust inside these environments, the main energy sources are the cosmic rays and soft X-rays. In this experiment, 150 eV photons (~ 4 x $10^{11}$ photons $cm^{-2}$ $s^{-1}$ ~ $10^3$ erg $cm^{-2}$ $s^{-1}$) were employed from the toroidal grating monocromator (TGM) beamline of LNLS (Pilling et al. 2011).

The search for amino acids, nucleobases, and related compounds in the interstellar medium/comets has been performed for the last 30 years. Recently, some traces (upper limits) of these molecules (e.g. glycine and pyrimidine) have been detected in molecular clouds, protostars, and comets. The search for these biomolecules in meteorites, on the contrary, has revealed an amazing number, up to several parts per million! This chemical dichotomy between meteorites and interstellar medium/comets remains a big puzzle in the field of astrochemistry and in the investigation about the origin of life.

The diluted samples were deposited onto a $CaF_2$ substrate by drop casting following solvent evaporation. The sample thicknesses were measured with a Dektak perfilometer and were of the order of 1-3 microns. The samples were placed into a vacuum chamber ($10^{-5}$ mbar) and exposed to different radiation doses up to 20 h. The experiments were performed at room temperature and took several weeks. *In-situ* sample analysis were performed by a Fourier transform infrared spectrometer (FTIR) coupled to the experimental vacuum chamber.

Figure 25a shows the FTIR spectra of amino acid DL-proline after different doses of 150-eV soft X-rays up to 7h. The FTIR spectra of RNA nucleobase uracil after different doses of 150-eV soft X-rays up to 17 h is showed in figure 25b. Figure 25c shows the integrated absorbance spectra of the solid-phase samples as a function of irradiation time (adapted from Pilling et al. 2011).

From the variation observed in the integrated area of IR spectrum (3000 to 900 $cm^{-1}$) along the different radiation doses, photodissociation rates and half-life were determined by employing Eq. 16 and Eq. 17, respectively. The results showed that amino acids can survive at least ~ $7 \times 10^5$ and ~$7 \times 10^8$ years in dense molecular clouds and protoplanetary disks, respectively. For the nucleobases, the photostability is even higher, being about 2-3 orders of magnitude higher than found for the most radiation sensitive amino acids (Pilling et al.

2011). This high degree of survivability could be attributed to the low photodissociation cross section of these molecules at soft X-ray, combined with the protection promoted by the dust which reduces the X-ray field inside denser regions. During planetary formation (and after), these molecules, trapped in and on dust grains, meteoroids, and comets, could be delivered to the planets/moons possibly allowing pre-biotic chemistry in environments where water was also found in liquid state.

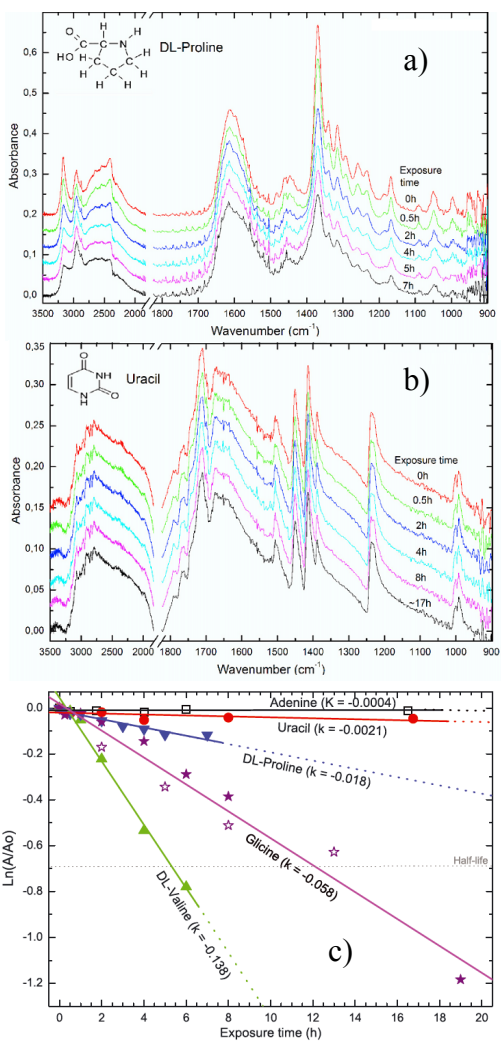

Fig. 25. a) FTIR spectra of amino acid DL-proline after different doses of 150-eV soft X-rays up to 7h. b) FTIR spectra of RNA nucleobase uracil after different doses of 150-eV soft X-rays up to 17 h. c) Integrated absorbance spectra of the solid-phase samples as a function of irradiation time. For each compound, the photodissociation rate (k) is also indicated (adapted from Pilling et al. 2011).

## 3. Conclusion

The presence of soft x-rays is very important for the chemical evolution of interstellar medium and other astrophysical environments close to young and bright stars. Soft X-rays can penetrate deep in molecular clouds and protostellar disks and trigger chemistry in regions in which UV stellar photons do not reach. The effects of soft X-rays in astrophysical ices are also remarkable because they release secondary electrons in and on the surface of the ices, which trigger a new set or chemical reactions. In this chapter, we presented several techniques employing soft X-rays in experimental simulation of astrophysical environments, which help us to understand the chemical evolution of these regions.

## 4. Acknowledgment

The authors would like to the thank the Brazilian agency FAPESP (#2011/14590-9) for the financial support. The authors also thank Ms. A. Rangel for the English revision of this book chapter.